\documentclass[english]{article}
\pdfoutput=1
\usepackage[latin1]{inputenc}
\usepackage{float}
\usepackage{amsmath}
\usepackage{graphicx}
\usepackage{amssymb}

\makeatletter

\usepackage{babel}
\makeatother

\begin{document}

\title{The optomechanical instability \\
in the quantum regime}

\author{Max Ludwig, Björn Kubala, and Florian Marquardt}

\maketitle
Department für Physik, Arnold Sommerfeld Center for Theoretical Physics,
and Center for NanoScience, Ludwig-Maximilians-Universität München,
Theresienstr. 37, D-80333 München, Germany

\begin{abstract}
We consider a generic optomechanical system, consisting of a driven
optical cavity and a movable mirror attached to a cantilever. Systems
of this kind (and analogues) have been realized in many recent experiments.
It is well known that those systems can exhibit an instability towards
a regime where the cantilever settles into self-sustained oscillations.
In this paper, we briefly review the classical theory of the optomechanical
instability, and then discuss the features arising in the quantum
regime. We solve numerically a full quantum master equation for the
coupled system, and use it to analyze the photon number, the cantilever's
mechanical energy, the phonon probability distribution and the mechanical
Wigner density, as a function of experimentally accessible control
parameters. We observe and discuss the quantum-to-classical transition
as a function of a suitable dimensionless quantum parameter.
\end{abstract}

\section{Introduction}

\newcommand{\prl}{Phys. Rev. Lett.}
\newcommand{\prb}{Phys. Rev. B}
\newcommand{\pra}{Phys. Rev. A}

\begin{figure}
\begin{centering}
\includegraphics[width=0.8\columnwidth]{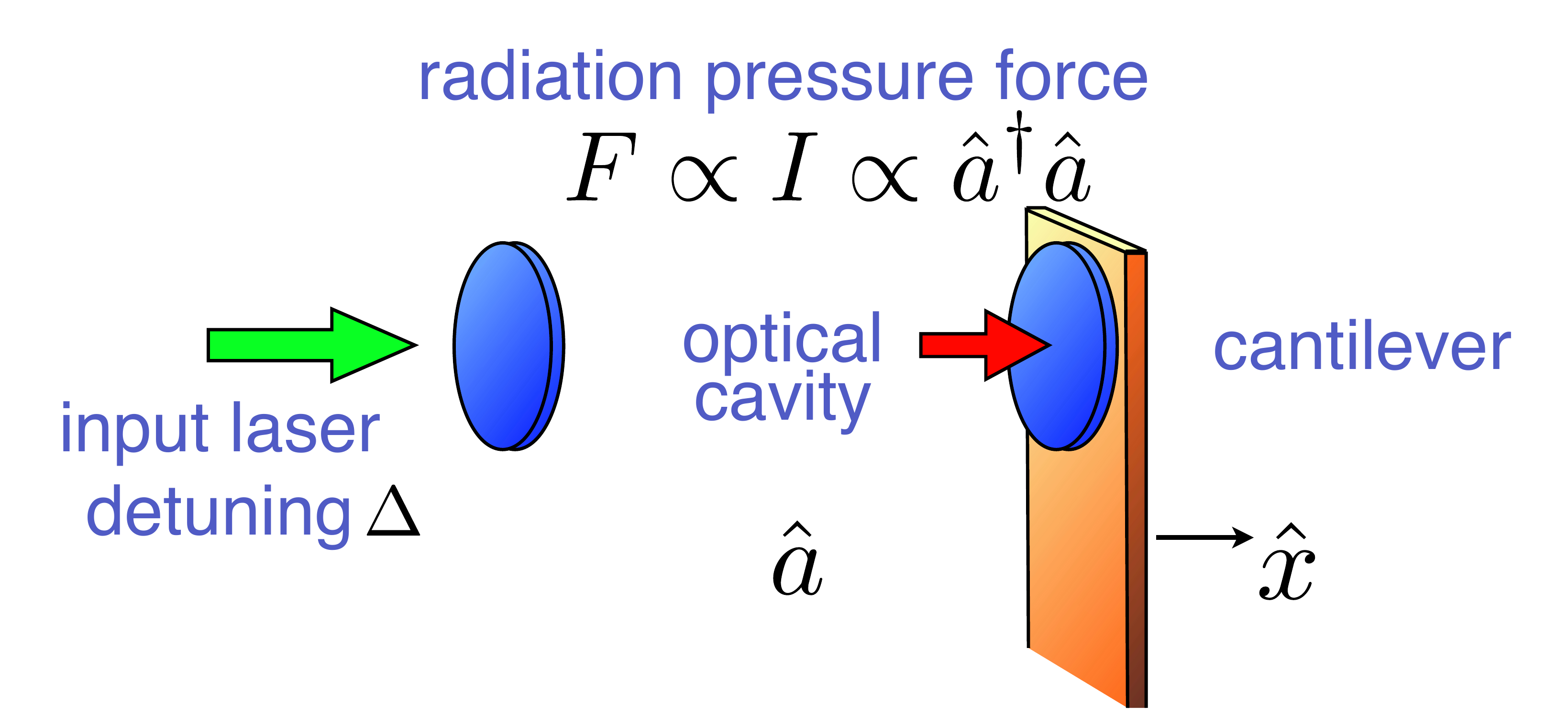}
\par\end{centering}

\caption{\label{figsetup}The basic optomechanical setup.}

\end{figure}

Light interacting with matter can not only be scattered, absorbed
and emitted by individual atoms, but it can also lead to mechanical
effects. The radiation pressure of light was first directly observed
in the seminal experiments of Nichols and Hull in 1901 and, independently,
by Lebedev, where it exerted a torque on a pair of glass mirrors inside
an evacuated chamber. Radiation pressure can also deflect the tail
of comets (as first hypothesized by Johannes Kepler), or change the
path of asteroids. The mechanical effects of light become most pronounced
in an optical cavity where the light intensity is resonantly enhanced,
and where one of the end-mirrors is made movable, e.g., by being attached
to a cantilever (Fig.~\ref{figsetup}). The pioneering theoretical
and experimental works in this domain are due to Braginsky \cite{1967_BraginskyManukin_PonderomotiveEffectsEMRadiation,1970_Braginsky_OpticalCoolingExperiment}.
In the 80s, strong effects were observed by the MPQ group of H.~Walther
in a setup using a macroscopic mirror \cite{1983_10_DorselWalther_BistabilityMirror}.
More recently, the trend has been to exploit the tools of microfabrication
to fabricate small cantilevers, nanobeams or other mechanical elements
that can be affected by light. The small masses, high mechanical quality
factors and (in some of the experiments) high optical finesse in these
setups increase the optomechanical effects. Several recent experiments
have demonstrated optomechanical cooling \cite{1999_10_Cohadon_CoolingMirrorFeedback,2004_12_HoehbergerKarrai_CoolingMicroleverNature,2006_07_Arcizet_CoolingMirror,2006_05_AspelmeyerZeilinger_SelfCoolingMirror,2006_11_Kippenberg_RadPressureCooling,2006_11_Bouwmeester_FeedbackCooling,2006_12_NergisMavalvala_LIGO,2007_07_Harris_MembraneInTheMiddle},
where the time-delayed light-induced forces lead to additional damping.
Such a scheme may ultimately be employed to cool down to the ground-state
of mechanical motion \cite{2007_01_Marquardt_CantileverCooling,2007_02_WilsonRae_Cooling}. 

On the other hand, the light-induced forces can also lead to a negative
contribution to the overall damping rate. At first, this increases
the mechanical $Q$ of the mechanical degree of freedom (hereafter
simply referred to as the cantilever), and can thus serve to amplify
the response to any noise source acting on the cantilever. Once the
overall {}``damping rate'' becomes negative, which can happen simply
by increasing the light power entering the cavity, we no longer have
damping but instead an instability \cite{1987_10_AguirregabiriaBel_DelayInducedInstabilityFP,1994_02_Fabre_SqueezingInCavityWithMovableMirror,2001_07_Braginsky_ParametricInstabilityFPCavity,2005_02_MarquardtHarrisGirvin_Cavity}.
The cantilever starts to oscillate at its eigenfrequency, with an
amplitude that at first increases exponentially and then saturates
to a steady-state value. These self-induced oscillations have by now
been observed in several experiments \cite{2004_KarraiConstanze_IEEE,2005_06_Vahala_SelfOscillationsCavity,2005_07_VahalaTheoryPRL,2007_11_LudwigNeuenhahn_SelfInducedOscillations}.
Theoretical studies predict an intricate attractor diagram \cite{2005_02_MarquardtHarrisGirvin_Cavity},
which can display multiple stable attractors (i.e. possible oscillation
amplitudes) for a given set of fixed external parameters. This attractor
diagram has recently been observed and studied systematically in a
low-finesse setup dominated by bolometric forces, which exhibited
the unexpected feature of simultaneous excitation of several mechanical
modes \cite{2007_11_LudwigNeuenhahn_SelfInducedOscillations}. 

Similar physics has by now been observed in a variety of other systems
which do not contain any optical elements. This includes driven LC
circuits coupled to cantilevers \cite{2007_Wineland_RFcircuitCooling}, single-electron
transistors and microwave cavities coupled to nanobeams \cite{PhysRevLett.93.136802,2005_11_Clerk_SSET_Cooling,2005_11_Blencowe_SET_NJP,2006_08_Schwab_CPB_Molasses,2006_BennettClerk_NEMSLaser,2007_Armour_ResonatorSSET,2007_Rodrigues_InstabilitySSET,2008_01_LehnertMicrowaveNanomechanics},
as well as clouds of cold atoms in an optical lattice inside a cavity
\cite{2006_03_MeiserMeystre_CoupledDynamicsAtomsAndCantileverCavity,2007_08_Murch_AtomsCavityHeating}. 

The common characteristic of all of these systems is that they contain
some driven resonant quantum system (optical or microwave cavity,
LC circuit, superconducting single-electron transistor), whose resonance
frequency depends on the motion of a mechanical degree of freedom
(cantilever, nanobeam, deformation of a microtoroidal optical resonator,
collective coordinate of a cloud of atoms). Their Hamiltonian thus
is typically of the form

\begin{equation}
\hat{H}=\hbar\omega_{R}(\hat{x})\,\hat{a}^{\dagger}\hat{a}+\hbar\omega_{M}\,\hat{c}^{\dagger}\hat{c}+\ldots,\label{eq:HamBasic}\end{equation}
where $\hat{a}$ is the photon annihilation operator for the driven
resonator, whose frequency depends on the coordinate $\hat{x}=x_{{\rm \text{ZPF}}}(\hat{c}+\hat{c}^{\dagger})$
of the mechanical oscillator with frequency $\omega_{M}$. Usually,
it is possible to a very good approximation to expand $\omega_{R}$
to linear order in $\hat{x}$, which is the case we will assume in
the following. The additional terms not displayed in Eq.~(\ref{eq:HamBasic})
then describe the driving, as well as the damping and the fluctuation
terms coupling to both oscillators (see below). 

Cooling to the mechanical ground-state will generate the opportunity
to observe a variety of quantum effects in such systems, including
{}``cat'' states \cite{1999_05_Bose_Cat}, entanglement \cite{2003_09_Marshall_QSuperposMirror,2005_12_Pinard_EntanglingAndSelfCooling}
and Fock state detection \cite{2007_07_Harris_MembraneInTheMiddle,2008_Buks_NDMFockstates}.
For a recent review of optomechanical systems, see \cite{2007_12_KippenbergVahala_ReviewOptomechanics},
and \cite{2008_03_MarquardtClerkGirvin_ReviewOptomechanics} for the
quantum noise approach to cooling.

One question that may be asked about the quantum regime of these devices,
that is now being approached experimentally, is how the instability
discussed above changes due to quantum effects. We will answer this
question partially in the present paper. We note that there have recently
been some discussions of the quantum dynamics for the related instability
in electronic systems in the literature \cite{2006_BennettClerk_NEMSLaser,2007_Armour_ResonatorSSET,2007_Rodrigues_InstabilitySSET,2007_Blanter_NEMS}.
The most important dimensionless parameter entering our analysis will
be the {}``quantum parameter''

\begin{equation}
\zeta\equiv\frac{x_{{\rm ZPF}}}{x_{{\rm FWHM}}},\end{equation}
which denotes the ratio between the mechanical zero-point fluctuation
amplitude (a quantum parameter, $\propto\sqrt{\hbar}$) and the width
of the optical resonance (a classical quantity), measured in terms
of displacement. After some rearrangement, one can see that this is
essentially the {}``granularity'' parameter employed in the discussion
of the cold-atom experiment \cite{2007_08_Murch_AtomsCavityHeating}.
There, it was introduced by considering the total momentum kick a
single photon would impart to the mechanical element, as it is reflected
multiple times before leaving the cavity. The granularity parameter
then would be derived from the ratio of this kick to the mechanical
momentum ground-state uncertainty. 

Increasing the quantum parameter $\zeta$ will enhance quantum effects
on the motion of the cantilever. These include the effects of the
photon shot noise, as well as the mechanical zero-point fluctuations.
The purpose of the present paper is to discuss these features in their
dependence on $\zeta$. We note that $\zeta$ is rather small in the
current optomechanical experiments (reaching up to about $\zeta\sim10^{-3}$
in \cite{2006_11_Bouwmeester_FeedbackCooling}). However, given the
large variety of analogous systems that are now being considered,
we feel that it is justified to illustrate some of the salient features
of the {}``quantum-to-classical crossover'' by also analysing the
regime $\zeta\sim1$.

The remainder of this paper is organized as follows: We will first
introduce the model Hamiltonian, and, in particular, discuss the full
set of dimensionless parameters that are needed in our analysis. We
then review the classical regime of self-induced oscillations. In
particular, we will discuss the attractor diagram for the {}``resolved
sideband regime'' $\omega_{M}\gg\kappa$, which has not been discussed
before in this context. This regime is currently of considerable interest,
as it is crucial for ground-state cooling \cite{2007_01_Marquardt_CantileverCooling,2007_02_WilsonRae_Cooling},
and $\omega_{M}/\kappa\sim20$ has recently been realized experimentally
\cite{2007_09_Kippenberg_SidebandCooling}. Then we will turn to the
full quantum model, that is first being discussed in terms of the
rate equations that can yield the behaviour below the instability
threshold. Afterwards, the full-blown quantum dynamics of self-induced
oscillations will be analyzed using a master equation approach applied
to the coupled system consisting of optical mode and cantilever. We
will illustrate how the average mechanical energy, as a function of
laser detuning, approaches the known classical result when the quantum
parameter is sent to zero. In our present analysis, we focus on the
steady-state and also assume a thermal bath temperature of $T=0$.
In real optomechanical experiments, one would presumably first cool
down using the light field and then observe the onset of nonlinear
dynamics as the detuning is varied. Investigation of these effects
would entail studying the complete nonequilibrium time-evolution.

\section{Model and parameters}

In this section we present the Hamiltonian of the coupled cavity-cantilever
system. A reduced set of dimensionless parameters determining the
dynamics of the coupled system is identified. In particular, we introduce
a quantum parameter, which is absent in descriptions of the classical
dynamics~\cite{2005_02_MarquardtHarrisGirvin_Cavity,2007_11_LudwigNeuenhahn_SelfInducedOscillations}
and governs the crossover from classical to quantum behavior of the
coupled dynamics of cavity and cantilever.

\subsection{Hamiltonian}

To describe a system of a mechanical cantilever coupled to a driven
cavity, we consider the Hamiltonian

\begin{equation}
\hat{H}=\hbar\,(-\Delta-g\,(\hat{c}+\hat{c}^{\dagger}))\,\hat{a}^{\dagger}\hat{a}\:+\;\hbar\omega_{M}\hat{c}^{\dagger}\hat{c}\:+\:\hbar\alpha_{L}(\,\hat{a}+\hat{a}^{\dagger}\,)\:+\:\hat{H}_{\kappa}\:+\:\hat{H}_{\Gamma}\,,\label{eq:1}\end{equation}
which is written in the rotating frame of the driving laser field
of frequency $\omega_{L}$, with an amplitude set by $\alpha_{L}$.
The laser is detuned by $\Delta=\omega_{L}-\omega_{\text{cav}}$ with
respect to the optical cavity mode, described by photon annihilation
and creation operators $\hat{a}$ and $\hat{a}^{\dagger}$, and a
photon number $\hat{n}_{{\rm {\rm cav}}}=\hat{a}^{\dagger}\hat{a}$.
The cantilever (or, in general, mechanical element) of frequency $\omega_{M}$
and mass $m$ has a phonon number $\hat{n}_{M}=\hat{c}^{\dagger}\hat{c}$,
and its displacement is given as $\hat{x}=x_{\text{ZPF}}(\hat{c}+\hat{c}^{\dagger})$,
with the ground state position uncertainty (mechanical zero-point
fluctuations) $x_{\text{ZPF}}=\sqrt{\hbar/(2m\omega_{M})}$. The optomechanical
coupling, between the optical field and the mechanical displacement,
is characterized by the parameter $g$. In the simplest case, with
a movable, fully reflecting mirror at one end of an optical cavity,
we have $g=\omega_{{\rm cav}}x_{{\rm ZPF}}/L$, and thus $g(\hat{c}+\hat{c}^{\dagger})=\omega_{{\rm cav}}\hat{x}/L$,
with a radiation pressure force equal to $\hat{F}_{\text{rad}}=\hat{a}^{\dagger}\hat{a}\,\hbar g/x_{\text{ZPF}}$.
The decay of a photon and the mechanical damping of the cantilever
are captured by $\hat{H}_{\kappa}$ and $\hat{H}_{\Gamma_{M}}$, respectively.
They describe coupling to a bath leading to a cavity damping rate
$\kappa$ and mechanical damping $\Gamma_{M}$. Note that each of
the parameters $\Delta,\, g,\,\omega_{M},\,\alpha_{L}$ has the dimension
of a frequency.

\subsection{Reduction to a set of dimensionless and independent parameters}

We now identify the dimensionless parameters the system dynamics depends
on. Expressed in terms of the mechanical oscillator frequency $\omega_{M}$,
the parameters describing the classical system are\begin{eqnarray*}
\text{mechanical damping} & : & \Gamma_{M}/\omega_{M}\\
\text{cavity decay} & : & \kappa/\omega_{M}\\
\text{detuning} & : & \Delta/\omega_{M}\\
\text{driving strength} & : & \mathcal{P}=8|\alpha_{L}|^{2}g^{2}/\omega_{M}^{4}=\omega_{{\rm cav}}\kappa^{2}E_{{\rm max}}^{{\rm cav}}/(\omega_{M}^{5}mL^{2}).\end{eqnarray*}
Here $E_{{\rm max}}^{{\rm cav}}$ is the light energy circulating
inside the cavity when the laser is in resonance with the optical
mode. The quantum mechanical nature of the system is described by
the {}``quantum parameter'' $ $$\zeta$, comparing the magnitude
of the cantilever's zero-point fluctuations, $x_{\text{ZPF}}$, with
the full width at half maximum (FWHM) of the cavity (translated into
a cantilever displacement $x_{\text{FWHM}}$) \[
\text{quantum parameter}\::\:\zeta=\frac{x_{\text{ZPF}}}{x_{\text{FWHM}}}=\frac{g}{\kappa}\,.\]
The resonance width of the cavity can be expressed as $x_{FWHM}=\kappa L/\omega_{\text{cav}}$,
where $L$ is the cavity's length. The quantum parameter $\zeta$
vanishes in the classical limit $\hbar\rightarrow0$, as the zero-point
fluctuations $x_{{\rm ZPF}}$ of the cantilever go to zero. The magnitude
of $\zeta$ determines the effect of quantum fluctuations on the dynamics
of the coupled cavity-cantilever system.

In later parts of this paper we will discuss the motion of the mechanical
cantilever due to the driving of the cavity, both in a classical and
a quantum mechanical picture. Such motion can be characterized by
its energy $E_{M}$, which in the classical case directly follows
from the oscillation amplitude $A$ of the cantilever: $E_{M,\text{cl}}=\frac{1}{2}m\omega_{M}^{2}A^{2}$.
In a quantum mechanical treatment, the energy is obtained from the
expectation value of the occupation number of the oscillator: $E_{M,\text{qm}}=\hbar\omega_{M}\langle\hat{n}_{M}\rangle$,
where we exclude the zero-point energy. The dimensionless ratio of
the cantilever energy $E_{M}$ to a characteristic classical energy
scale of the system is then easily compared for the two approaches.
To set this characteristic energy scale, we take the energy $E_{0}=\frac{1}{2}m\omega_{M}^{2}x_{\text{FWHM}}^{2}$
associated with an oscillation amplitude $x_{\text{FWHM}}$ of the
mechanical cantilever which moves the cavity just out of its resonance.
Note that $E_{M}/E_{0}=(A/x_{{\rm FWHM}})^{2}$ in the classical case,
and $E_{M}/E_{0}=4\zeta^{2}\left\langle \hat{n}_{M}\right\rangle $
in the quantum version.

\section{Dynamics of the system}

In this section, we will first briefly recapitulate the results of
a classical treatment of the optomechanical system \cite{2005_02_MarquardtHarrisGirvin_Cavity,2007_11_LudwigNeuenhahn_SelfInducedOscillations}.
We will concentrate on the regime of blue-detuned excitation of the
cavity, where the cantilever motion is amplified and self-induced
oscillations can occur. Amplification behaviour of the coupled system
(away from the regime of self-induced oscillations) can be understood
within a simple rate equation approach, which captures the effect
of photon shot noise leading to fluctuations of the radiation pressure
force acting on the cantilever \cite{2007_01_Marquardt_CantileverCooling}.
The full quantum mechanical treatment, employing the numerical solution
of a quantum master equation, can describe the crossover from heating/amplification
to classical self-induced oscillations of the coupled system, where
the quantum parameter $\zeta=x_{\text{ZPF}}/x_{\text{FWHM}}$ governs
the quantum-to-classical transition.

\subsection{Classical solution }

Heisenberg equations of motion for the cavity operator $\hat{a}$
and the cantilever position operator $\hat{x}$ can easily be derived
from the Hamiltonian, Eq.~\ref{eq:1}. To investigate the purely
classical dynamics of the coupled cavity-cantilever system, we replace
the operator $\hat{a}(t)$ by the complex light amplitude $\alpha(t)$
and the position operator of the cantilever $\hat{x}$ by its classical
counterpart. We thus arrive at:

\begin{equation}
\dot{\alpha}=[i(\Delta+g\frac{x}{x_{\text{ZPF}}})-\frac{\kappa}{2}]\,\alpha-i\alpha_{L}\label{eq:2}\end{equation}

\begin{equation}
\ddot{x}=-\omega_{M}^{2}x+\frac{\hbar g}{mx_{\text{ZPF}}}\left|\alpha\right|^{2}-\Gamma_{M}\dot{x}\,.\label{eq:3}\end{equation}
Here fluctuations (both the photon shot noise as well as intrinsic
mechanical thermal fluctuations) have been neglected, to obtain the
purely deterministic classical solution. The variables $t$, $x$
and $\alpha$ can be rescaled \cite{2005_02_MarquardtHarrisGirvin_Cavity}
as $\tilde{t}=\omega_{M}t;\:\tilde{\alpha}=i\alpha\omega_{M}/(2\alpha_{L});\;\tilde{x}=gx/(\omega_{M}x_{\text{ZPF}})\,$,
so that the coupled equations of motion contain only the dimensionless
parameters ${\cal P},\,\Delta/\omega_{M},\,\kappa/\omega_{M},\,$
and $\Gamma_{M}/\omega_{M}$:

\begin{eqnarray*}
\frac{d\tilde{\alpha}}{d\tilde{t}} & = & [i(\frac{\Delta}{\omega_{M}}+\tilde{x})-\frac{1}{2}\frac{\kappa}{\omega_{M}}]\tilde{\alpha}+\frac{1}{2}\\
\frac{d^{2}\tilde{x}}{d\tilde{t}^{2}} & = & -\tilde{x}+\mathcal{P}\left|\tilde{\alpha}\right|^{2}-\frac{\Gamma_{M}}{\omega_{M}}\frac{d\tilde{x}}{d\tilde{t}}\,.\end{eqnarray*}
Crucially, the quantum parameter $\zeta$ cannot and does not feature
in these equations.

Apart from a static solution $x(t)\equiv{\rm const}$, this system
of coupled differential equations can show self-induced oscillations.
In such solutions, the cantilever conducts an approximately sinusoidal
oscillation at its unperturbed frequency, $x(t)\approx\bar{x}+A\cos(\omega_{M}t)$.
The radiation pressure affects the cantilever motion rather weakly,
so that the oscillation amplitude $A$ varies only slowly and can
be taken as constant during one oscillation period. The light amplitude
then shows the dynamics of a damped, driven oscillator, which is swept
through its resonance, see Eq.~\eqref{eq:2}; an exact solution for
the light amplitude $\alpha(t)$ can be given as a Fourier series
containing harmonics of the cantilever frequency $\omega_{M}$ \cite{2005_02_MarquardtHarrisGirvin_Cavity}:

\begin{equation}
\left|\tilde{\alpha}(\tilde{t})\right|=\left|\sum_{n}\tilde{\alpha}_{n}e^{in\tilde{t}}\right|,\end{equation}
with

\begin{equation}
\tilde{\alpha}_{n}=\frac{1}{2}\frac{J_{n}(-\tilde{A})}{in+\kappa/(2\omega_{M})-i(\bar{\tilde{x}}+\Delta/\omega_{M})}\,.\end{equation}

The dependence of oscillation amplitude, $A$, and average cantilever
position, $\bar{x}$, on the dimensionless system parameters can be
found by two balance conditions: Firstly, the total force on the cantilever
has to vanish on average, and, secondly, the power input into the
mechanical oscillator by the radiation pressure on average has to
equal the friction loss. 

The force balance condition determines the average position of the
oscillator, yielding an implicit equation for $\bar{x}$, \begin{equation}
\langle\ddot{x}\rangle\equiv0\quad\Leftrightarrow\quad m\omega_{M}^{2}\bar{x}=\left\langle F_{{\rm rad}}\right\rangle =\frac{\hbar g}{mx_{\text{ZPF}}}\langle\left|\alpha(t)\right|^{2}\rangle\,,\label{eq:forcebalance}\end{equation}
where the average radiation force, $\langle F_{\text{rad}}\rangle$
is a function of the parameters $\bar{x}$ and $A$.

\begin{figure}[H]
\begin{centering}
\includegraphics[width=0.95\columnwidth]{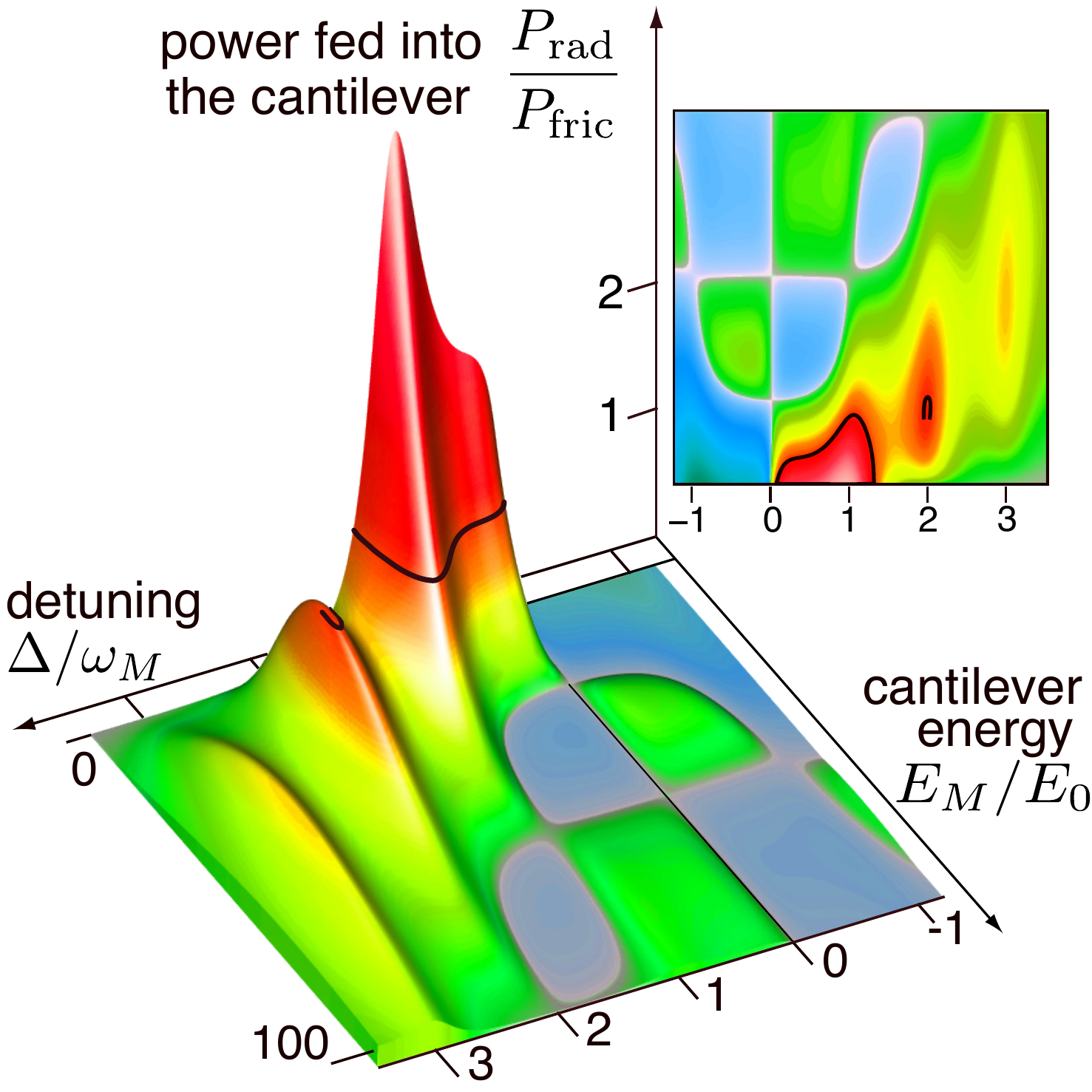}\caption{Classical self-induced oscillations of the coupled cavity-cantilever
system. The radiation pressure acting on the cantilever provides an
average mechanical power input of $P_{{\rm rad}}$. The ratio $P_{{\rm rad}}/P_{{\rm fric}}$
of this power $P_{{\rm rad}}$ vs. the loss due to mechanical friction,
$P_{{\rm fric}}$, is shown as a function of the detuning $\Delta$
and the cantilever's oscillation energy $E_{M}$, at fixed laser input
power $\mathcal{P}$. The oscillation energy $E_{M}=m\omega_{M}^{2}A^{2}/2$
is shown in units of $E_{0}$, where $E_{M}/E_{0}=(A/x_{{\rm FWHM}})^{2}$.
Self-induced oscillations require $P_{{\rm rad}}=P_{{\rm fric}}$.
This condition is fulfilled along the horizontal cut at $P_{{\rm rad}}/P_{{\rm fric}}=1$
(see black line and the inset depicting the same plot, viewed from
above). These solutions are stable if the ratio $P_{{\rm rad}}/P_{{\rm fric}}$
decreases with increasing oscillation amplitude A. The blue regions
at the floor of the plot indicate that $P_{{\rm rad}}$ is negative,
resulting in cooling. The cavity decay rate is $\kappa=0.5\omega_{M}$,
the mechanical damping is chosen as $\Gamma_{M}/\omega_{M}=1.47\cdot10^{-3}$,
and the input power as ${\cal P}=6.05\cdot10^{-3}\,$; these parameters
are also used in figures~\ref{fig:2}, \ref{fig:3}, \ref{fig:4},
and \ref{fig:5}, and will be referred to as $\Gamma_{M}^{*}$ and
$\mathcal{P}^{*}$. \label{fig:1}}

\par\end{centering}
\end{figure}

The balance between the mechanical power gain due to the light-induced
force, $P_{{\rm rad}}=\left\langle F_{{\rm rad}}\dot{x}\right\rangle $,
and the frictional loss $P_{{\rm fric}}=\Gamma_{M}\left\langle \dot{x}^{2}\right\rangle $
follows from \begin{equation}
\langle\dot{x}\ddot{x}\rangle\equiv0\quad\Leftrightarrow\quad\langle F_{\text{rad}}\dot{x}\rangle=\Gamma_{M}\langle\dot{x}^{2}\rangle.\label{eq:powerbalance}\end{equation}
For each value of the oscillation amplitude $A$ we can now plot the
ratio between radiation power input and friction loss, $P_{{\rm rad}}/P_{{\rm fric}}=\langle F_{\text{rad}}\dot{x}\rangle/(\Gamma_{M}\langle\dot{x}^{2}\rangle)$,
after eliminating $\bar{x}$ using Eq.~\ref{eq:forcebalance}. This
is shown in Fig.~\ref{fig:1}. Power balance is fulfilled if this
ratio is one, corresponding to the contour line $P_{{\rm rad}}/P_{{\rm fric}}=1$.
If the power input into the cantilever by radiation pressure is larger
than frictional losses (i.e., for a ratio larger than one), the amplitude
of oscillations will increase, otherwise it will decrease. Stable
solutions (dynamical attractors) are therefore given by that part
of the contour line where the ratio decreases with increasing oscillation
amplitude (energy), as shown in Fig.~\ref{fig:1}. 

Changing the (dimensionless) mechanical damping rate $\Gamma_{M}/\omega_{M}$
will scale the plot in Fig.~\ref{fig:1} along the vertical axis,
so that the horizontal cut at one yields a different contour line
of stable solutions {[}a changed input power $\mathcal{P}$ gives
a similar scaling, but leads to further changes in the solution, as
$\mathcal{P}$ also enters the force balance condition, Eq.~\eqref{eq:forcebalance}].
Decreasing mechanical damping or increasing the power input will increase
the plot height in Fig.~\ref{fig:1}, so that the amplitude/energy
of oscillation of the stable solution increases.

\begin{figure}[H]
\begin{centering}
\includegraphics[width=0.95\columnwidth]{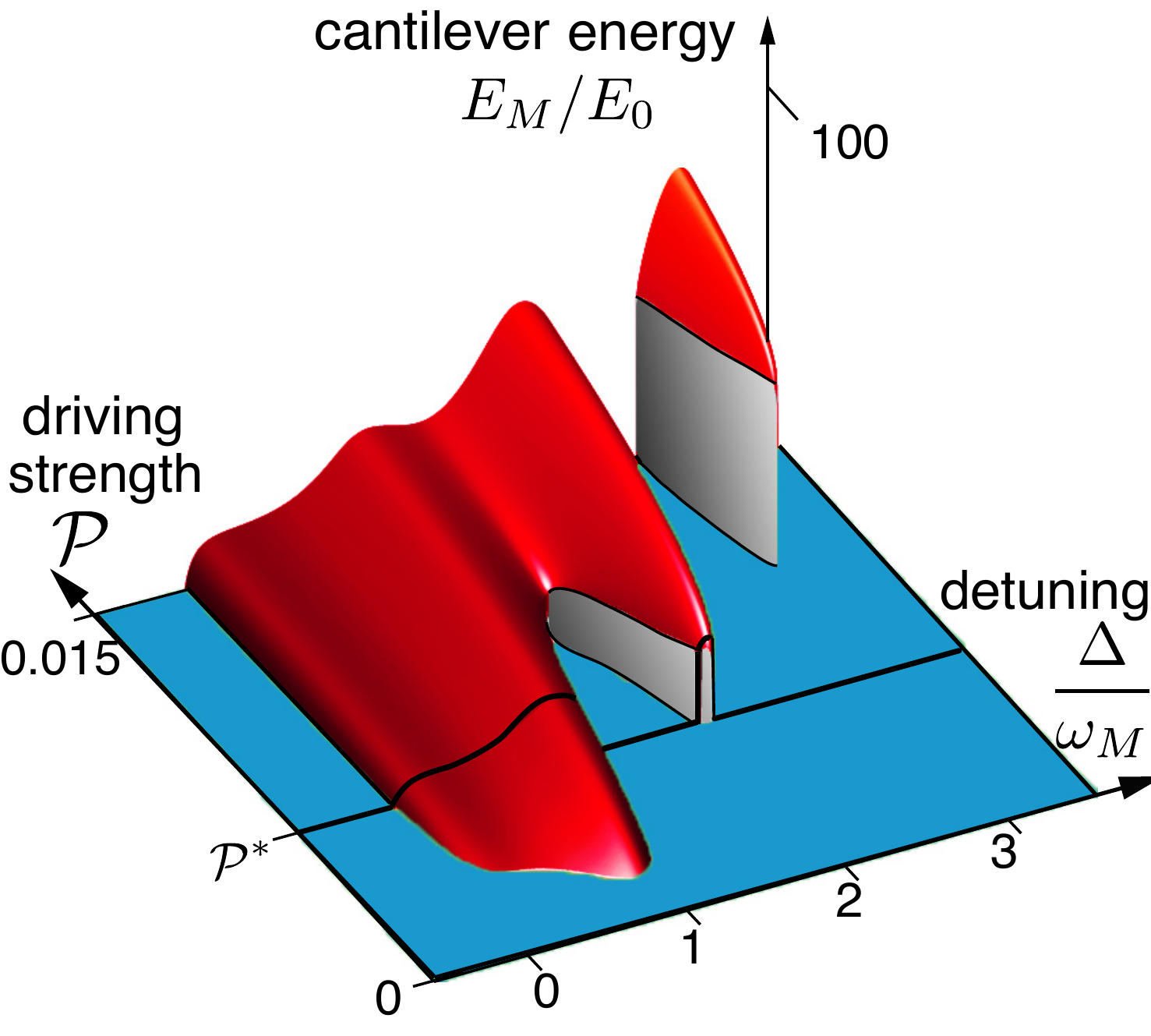}\caption{Cantilever oscillation energy $E_{M}\propto A^{2}$ versus detuning
$\Delta$ and laser input power $\mathcal{P}$. This plot (in contrast
to Fig.~\ref{fig:1}) shows only the stable oscilation amplitude,
but as a function of variable input power. The particular value $\mathcal{P}^{*}$
corresponding to Fig.~\ref{fig:1}, and the resulting profile of
oscillation amplitudes are indicated by a black line. The green floor
of the plot indicates regions without self-induced oscillations. The
other system parameters are as in Fig.~\ref{fig:1}. The continuous
onset of the self-oscillations in the sidebands at $\Delta/\omega_{M}=0,\,1$
(which merge for the present parameter values) represents a supercritical
Hopf bifurcation, from $A=0$ to $A\neq0$. At higher sidebands, an
attractor with a finite $A\neq0$ appears discontinuously, while $A=0$
remains a stable solution. \label{fig:2}}

\par\end{centering}
\end{figure}

While the surface or contour plots in Fig.~\ref{fig:1} allow a discussion
of general features of the self-induced oscillations, such as the
multistabilities discussed in Ref.~\cite{2005_02_MarquardtHarrisGirvin_Cavity},
a slightly different representation of the classical solution is more
amenable to an easier understanding of the particular dynamics of
the system for a certain set of fixed system parameters. Figure~\ref{fig:2}
shows the cantilever energy $E_{M,\text{cl}}=\frac{1}{2}m\omega_{M}^{2}A^{2}$
in terms of the classical energy scale $E_{0}=\frac{1}{2}m\omega_{M}^{2}x_{\text{FWHM}}^{2}$
as function of driving ${\cal {P}}$ and detuning $\Delta/\omega_{M}$.
These are the parameters that can typically be varied in a given experimental
setup. 

For sufficiently strong driving, self-induced oscillations appear
around integer multiples of the cantilever frequency, $\Delta\approx n\omega_{M}$.
For a cavity decay rate $\kappa=0.5\omega_{M}$ assumed in Fig.~\ref{fig:2},
the different bands are distinguishable at lower driving; for larger
$\kappa$ (or for stronger driving), the various `sidebands' merge.
For the lower-order sidebands, the nonzero amplitude solution connects
continuously to the zero amplitude solution, which becomes unstable.
This is an example of a (supercritical) Hopf bifurcation into a limit
cycle.

The vertical faces, shown gray in Fig.~\ref{fig:2}, for $\Delta\approx2\omega_{M}$
and $\Delta\approx3\omega_{M}$ are connected to the sudden appearance
of attractors with a finite amplitude. For example, while approaching
the detuning of $\Delta=2\omega_{M}$ at fixed $\mathcal{P}$ (the
solid line in Fig.~\ref{fig:2} refers to ${\cal P}=1.47\cdot10^{-3}$),
a finite amplitude solution appears, although $A=0$ remains stable.
In Ref.~\cite{2005_02_MarquardtHarrisGirvin_Cavity} the existence
of higher-amplitude stable attractors and, correspondingly, dynamic
multistability were discussed.

\subsection{Rate equation approach}

Before embarking on a full quantum-mechanical treatment of the coupled
cavity-cantilever system, it is instructive to discuss a more simple
method to capture some nonclassical effects, in particular the response
of the cantilever to the photon shot noise. For that purpose, we consider
the shot noise spectrum of the driven cavity, decoupled from the cantilever,
\begin{equation}
S_{FF}(\omega)=\left(\frac{\hbar g}{x_{\text{ZPF}}}\right)^{2}S_{nn}(\omega)=\left(\frac{\hbar g}{x_{\text{ZPF}}}\right)^{2}\bar{n}\frac{\kappa}{(\omega+\Delta)^{2}+(\kappa/2)^{2}}\,,\label{eq:S_FF}\end{equation}
where\begin{equation}
\bar{n}=\frac{{\cal P}}{8\zeta^{2}}\frac{(\omega_{M}/\kappa)^{2}}{(\Delta/\omega_{M})^{2}+(\kappa/2\omega_{M})^{2}}\label{eq:nav}\end{equation}
is the mean number of photons in the cavity. The maximum occupation
$n_{\text{max}}=\mathcal{P}\omega_{M}^{4}/(2\kappa^{4}\zeta^{2})=4\alpha_{L}^{2}/\kappa^{2}$
occurs at zero detuning. We note that in using the unperturbed, intrinsic
shot noise spectrum for an optical cavity in the absence of optomechanical
effects, we neglect the modification of that spectrum due to the backaction
of the cantilever motion.

The asymmetry of the shot noise spectrum is important for the dynamics
of the cantilever. The spectral density of the radiation-pressure
force at positive frequency $\omega_{M}$ (negative frequency $-\omega_{M}$)
yields the probability of the cavity absorbing a phonon from (emitting
a phonon into) the cantilever~\cite{2007_01_Marquardt_CantileverCooling}.

For a red-detuned laser impinging on the cavity ($\Delta<0$), the
cavity's noise spectrum peaks at positive frequencies and the cavity
tends to rather absorb energy from the cantilever. As a consequence,
the mechanical damping rate for the cantilever is increased, leading
to cooling if one starts with a sufficiently hot cantilever. In the
opposite Raman-like process taking place at $\Delta>0$, a blue-detuned
laser beam will preferentially lose energy to the cantilever, so that
it matches the cavity's resonance frequency. The effective optomechanical
damping rate, 

\begin{equation}
\Gamma_{\text{opt}}=\zeta^{2}\kappa^{2}[S_{nn}(+\omega_{M})-S_{nn}(-\omega_{M})]\,,\label{eq:gopt}\end{equation}
is then negative. The corresponding heating of the mechanical cantilever
is counteracted by the mechanical damping $\Gamma_{M}$ . Simple rate
equations for the occupancy of the cantilever yield a thermal distribution
for the cantilever phonon occupation number $n_{M}$, with \cite{2007_01_Marquardt_CantileverCooling}\begin{equation}
\langle\hat{c}^{\dagger}\hat{c}\rangle=\langle\hat{n}_{M}\rangle=\frac{\zeta^{2}\kappa^{2}S_{nn}(-\omega_{M})+\bar{n}_{{\rm th}}\Gamma_{M}}{\Gamma_{{\rm opt}}+\Gamma_{M}}\,.\label{eq:N_c_rate_equations}\end{equation}
The effective temperature, $T_{\text{eff}}$, is related by $\langle\hat{n}_{M}+1\rangle/\langle\hat{n}_{M}\rangle=\exp[\hbar\omega_{M}/(k_{B}T_{\text{eff}})]$
to the mean occupation number. The equilibrium mechanical mode occupation
number, $\bar{n}_{{\rm th}}$, is determined by the mechanical bath
temperature, which is taken as zero in the following. In contrast
to first appearance, the mean occupation number of the cantilever
given in Eq.~(\ref{eq:N_c_rate_equations}) does not depend on the
quantum parameter $\zeta$, as $\zeta^{2}S_{nn}$ is independent of
$\zeta$. This is because $S_{nn}\sim\bar{n}\sim1/\zeta^{2}$, see
Eq.~(\ref{eq:nav}). The cantilever energy, therefore, only trivially
depends on the quantum parameter as $E_{M}/E_{0}=4\zeta^{2}\left\langle \hat{n}_{M}\right\rangle $,
so that it vanishes in the classical limit, where $\zeta^{2}\propto\hbar\rightarrow0$.

In general, the phonon number in Eq.~(\ref{eq:N_c_rate_equations})
can increase due to two distinct physical effects: On the one hand,
the numerator can become larger, due to the influence of photon shot
noise impinging on the cantilever, represented by $S_{nn}$. On the
other hand, the denominator can become smaller due to $\Gamma_{{\rm opt}}$
becoming negative. In the latter case, the fluctuations acting on
the cantilever (both thermal and shot noise) are amplified. This effect
is particularly pronounced just below the threshold of instability,
where $\Gamma_{M}+\Gamma_{{\rm opt}}=0$ (see below). 

\begin{figure}[H]
\begin{centering}
\includegraphics[width=0.95\columnwidth]{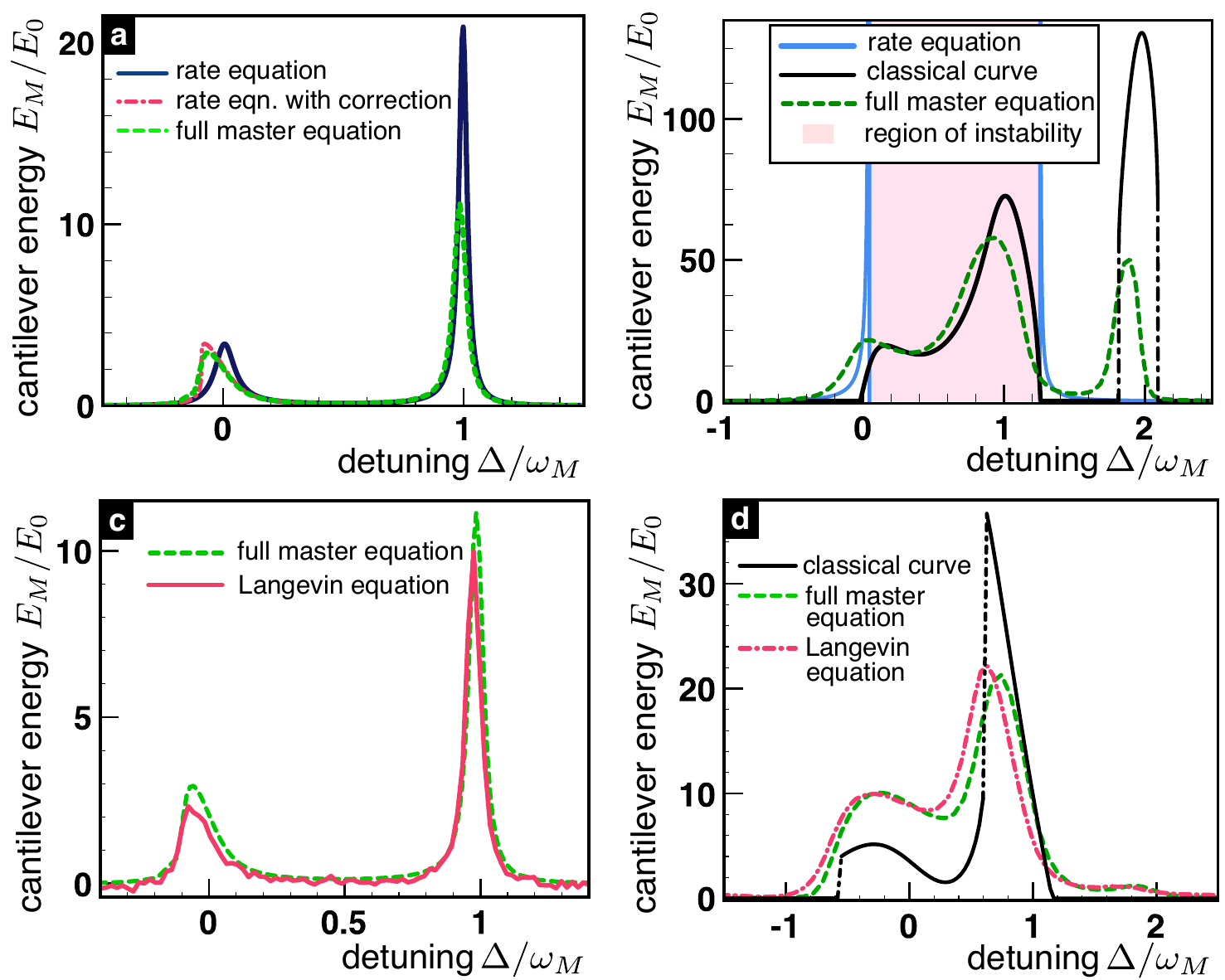}
\par\end{centering}

\caption{Cantilever energy versus detuning for a cavity driven below {[}(a),(c)]
and above {[}(b), (d)] the onset of self-induced oscillations. Note
$E_{M}/E_{0}=4\zeta^{2}\left\langle \hat{n}_{M}\right\rangle $. (a)
Below the onset, the cantilever amplitude would vanish according to
the classical analysis that does not incorporate fluctuations. However,
the cantilever is susceptible to the photon shot noise (the parameters
are $\kappa/\omega_{M}=0.1$, $\mathcal{P}=8.4\cdot10^{-3}$ , $\Gamma_{M}/\omega_{M}=5\cdot10^{-3}$,
and $\zeta=1.0$), leading to finite phonon numbers in the cantilever,
particularly around the resonance $\Delta=0$ and at the first sideband
$\Delta=\omega_{M}$ (see main text). This is captured by the full
quantum master equation, as well as (approximately) by the rate equation,
whose results improve when taking into account the corrections due
to the shift of the cantilever position $\bar{x}$. (b) For stronger
driving, the classical solution yields self-oscillations (the parameters
are ${\cal \mathcal{P}}^{*},\,\Gamma_{M}^{*}$ as in Fig.~\ref{fig:2},
but $\kappa/\omega_{M}=0.3$). The rate equation correctly predicts
the onset of the linear instability, but not the nonlinear regime.
{[}The shift in $\bar{x}$ was not taken into account, hence the slight
discrepancy vs. the classical solution] The master equation results
are shifted to lower detuning and describe sub-threshold amplification
and heating as well as self-induced oscillations above threshold,
modified and smeared due to quantum effects (as shown for a quantum
parameter of $\zeta=x_{\text{ZPF}}/x_{\text{FWHM}}=1$). (c) Including
the zero-point fluctuations in a semiclassical approach via Langevin
equations gives results that agree well with both the results from
the rate equation and the full master equation, shown here for parameters
as in (a). (d) Above the onset of self-induced oscillations the semiclassical
approach mimicks results from the quantum master equation partially.
The parameters for this plot are $\kappa/\omega_{M}=0.3$, $\Gamma_{M}=50\Gamma_{M}^{*}$,
$\mathcal{P}=20\mathcal{P}^{*}$, $\zeta=1$. \label{fig:3}}

\end{figure}

In the resolved sideband limit $\kappa\ll\omega_{M}$ (at weak driving)
the cantilever occupation $\langle\hat{n}_{M}\rangle$ will peak around
zero detuning, where the number of photons in the cavity is large,
and around a detuning of $\Delta=\omega_{M}$. At the latter value
of detuning the aforementioned Raman process is maximally efficient
as a photon entering the cavity will exactly match the resonance frequency
after exciting a phonon in the cantilever. This dependence of cantilever
occupation (or the corresponding energy) on the detuning is shown
in Fig.~\ref{fig:3}. 

The approach sketched above can be modified slightly to take account
of the modification of the cavity length due to a static shift of
the cantilever mirror by radiation pressure. Approaching the resonance
of the cavity from below, the increasing number of photons inside
the cavity will increase the cavity length due to their radiation
pressure on the mirror, bringing the system even closer to the resonance.
The effect of the static shift of the mirror on the mean occupation
of the cavity can be included self-consistently, leading to the tilt
of the peak around the resonance, shown by the dash-dotted line in
Fig.~\ref{fig:3}(a). The same figure also includes results of the
full quantum mechanical approach, which will be discussed in the next
section.

For larger $\kappa$, the two peaks in the cantilever excitation merge.
Higher-order sidebands are not resolved within this approach, since
they would require taking care of the modification of $S_{FF}$ due
to the cantilever's motion. 

Classical self-induced oscillations occur in a regime of larger driving,
where the optomechanical damping rate $\Gamma_{{\rm opt}}$ of Eq.~(\ref{eq:gopt})
becomes negative. They appear once amplification exceeds intrinsic
damping, i.e. when $\Gamma_{{\rm opt}}+\Gamma_{M}<0$. The simple
rate equation approach lacks any feedback mechanism to stop the divergence
of the phonon number. The classical solution demonstrates how this
feedback (i.e. the resulting change in the dynamics of the radiation
field) makes the mechanical oscillation amplitude saturate at a finite
level. In addition, it shows the onset of self-induced oscillations
to occur at a smaller detuning, due to the effective shift of the
cantilever position explained above.

In Fig.~\ref{fig:3}(b) we show results for the detuning dependence
of the mean energy of the cantilever obtained from this rate equation
approach below the threshold of classical self-induced oscillations.
The coupled cavity-cantilever system acts as an amplifier of fluctuations,
increasing the occupation of higher number states of the cantilever
well before classical oscillations set in. At the onset of classical
self-induced oscillations the rate equation result diverges. A full
quantum-mechanical treatment describes the crossover of the cantilever
dynamics from quantum-fluctuation induced heating to self-induced
oscillations as will be discussed now.

\subsection{Quantum master equation method}

The evolution of the coupled quantum system comprised of the cantilever
and the optical cavity is described by the Hamiltonian of Eq.~\eqref{eq:1}.
Dissipation arises from the coupling of the mechanical mode to a bath
and due to the opening of the cavity to the outside. While the former
results in mechanical damping with a rate $\Gamma_{M}$, the latter
is associated with the ringdown rate of the cavity $\kappa$. In the
present paper, we will assume the mechanical bath to be at zero temperature,
where quantum effects are most pronounced in steady state. A future,
more realistic treatment, should relax this assumption and deal with
the nonequilibrium dynamics that results when a mechanical system
is first cooled optomechanically and then switched to the unstable
side.

The system can be described by a reduced density matrix $\hat{\rho}$
for the mechanical cantilever mode and the optical mode of the cavity.
In the frame rotating at the laser frequency, the time evolution of
the density matrix $ $$\hat{\rho}$ is given by 

\begin{equation}
\frac{d}{dt}\hat{\rho}=\frac{[\hat{H}_{0},\hat{\rho}]}{i\hbar}\,+\,\Gamma_{M}\,\mathcal{D}[\hat{c}]\,+\,\kappa\,\mathcal{D}[\hat{a}]\,,\quad(T\equiv0)\label{eq:Lindblad}\end{equation}
where $\mathcal{D}[\hat{A}]=\hat{A}\hat{\rho}\hat{A}^{\dagger}\,-\,\frac{1}{2}\hat{A}^{\dagger}\hat{A}\hat{\rho}\,-\,\frac{1}{2}\hat{\rho}\hat{A}^{\dagger}\hat{A}$
denotes the standard Lindblad operator. The Hamilton operator $\hat{H}_{0}$
describes the coherent part of the evolution of the coupled cavity-cantilever
system, \[
\hat{H}=\hat{H}_{0}\:+\:\hat{H}_{\kappa}\:+\:\hat{H}_{\Gamma}\,.\]
For the numerical evaluation, we rewrite Eq.~\ref{eq:Lindblad} as
$d\hat{\rho}/dt=\mathcal{L}\hat{\rho}$, with a Liouvillian superoperator
$\mathcal{L}$. We then interpret the density matrix as a vector,
whose time evolution is governed by the matrix $\mathcal{L}$. The
density matrix at long times (in steady state) is then given by the
eigenvector of $\mathcal{L}$ with eigenvalue $0$. The numerical
calculation of this eigenvector is much more efficient than a simulation
of the full time evolution. Since we are dealing with large sparse
matrices, it is convenient to employ an Arnoldi method that finds
a few eigenvalues and eigenvectors of $\mathcal{L}$ by iterative
projection. For Hermitean matrices, the Arnoldi method is also known
as the Lanczos algorithm.

In practice, the numerical approach used here sets strong limits on
the dimension of the Hilbert space. We need to take into account the
$N_{a}$ lowest Fock states of the cavity and the $N_{c}$ lowest
Fock states of the mechanical cantilever, resulting in a Liouvillian
superoperator with $(N_{a}\times N_{c})^{4}$ elements. This puts
more severe restrictions on our treatment of the coupled cavity-cantilver
system than encountered in similar treatments of comparable systems.
For example, nanoelectromechanical systems, where an oscillator is
coupled to a normal-state or superconducting single-electron transistor
(SET), will have to account for only a very limited number of charge
states of the SET (namely those few involved in the relevant transport
cycle). As a consequence, a larger number of Fock states can be included,
e.g., $70$ number states of the oscillator were kept in Ref.~ \cite{2007_Armour_ResonatorSSET}.
In some cases it was furthermore considered sufficient to treat only
the incoherent dynamics of the mechanical oscillator, i.e., only the
elements of the density matrix diagonal in the oscillator's Fock space,
thereby reaching $200$ number states of a mechanical mode coupled
to a normal-state SET \cite{2008_MerloHauptCavaliereSassetti_NEMSubpoissonian}.
The restricted number of Fock states that can be considered here makes
it more difficult to fully bridge the gulf to the classical regime
of motion of the mechanical cantilever. {[}$(N_{a},N_{c})=$ (8,16)
for Fig.~\ref{fig:3}(a),(c),(d), (4,22) for Figs.~\ref{fig:3}(b),~\ref{fig:4}
and for the first two panels of \ref{fig:5}, (3,35) for the last
panel of Fig. \ref{fig:5}]

A first comparison of results of the quantum master equation to the
classical solution and the results of the rate equation was already
shown in Fig.~\ref{fig:3}. We find that the full quantum results
do not qualitatively differ from the rate equation results provided
the parameters are chosen sufficiently far from the onset of self-induced
oscillations. However, only the quantum master equation approach is
able to describe the crossover from sub-threshold quantum fluctuations
(where $E_{M}\propto\zeta^{2}$) to the large classical cantilever
energies associated with self-induced oscillations. 

In Fig.\ \ref{fig:4} we demonstrate the influence of the quantum
parameter $\zeta=x_{\text{ZPF}}/x_{\text{FWHM}}$ governing the crossover
between the classical and the quantum regime. 

Figure~\ref{fig:4}(a) shows the cavity photon number, normalized
to its value at resonance, $n_{\text{max}}$. For our choice of driving
parameter $\mathcal{P}$, the maximal occupation $n_{\text{max}}$
is low, so that a small number of Fock states suffices for describing
the cavity in the quantum master equation. This allows to account
for enough number states of the cantilever to reach the regime of
self-induced oscillations. The classical solution (solid black line)
consists of the broad Lorentzian of the isolated cavity, on top of
which additional peaks appear. These are due to the classical self-induced
oscillations occuring at the sidebands $\Delta=\omega_{M},\,2\omega_{M},\hdots$
in the coupled cavity-cantilever system. Figure~\ref{fig:4}(c) displays
the cantilever energy $E_{M}/E_{0}$ as a function of the detuning,
$\Delta/\omega_{M}$, with features that parallel those found for
the photon number. The classical curve in (b), shown in black, corresponds
to the cut indicated by the solid line in Fig.~\ref{fig:2}. For
the chosen driving power, the second sideband at $\Delta=\omega_{M}$
just starts to appear, while the first sideband is merged with the
resonance at $\Delta=0$, which shows up as a slight shoulder. The
sharpness and strength of these features also depend on the values
of mechanical damping and cavity decay rate. Results of our solution
of the quantum master equation are shown for three different values
of the quantum parameter $\zeta=x_{\text{ZPF}}/x_{\text{FWHM}}$.
Due to restrictions of the numerical resources, it was not feasible
to map out a wider range of values of the parameter $\zeta$, although
the range analysed here already suffices to describe the quantum-classical
crossover. 

The quantum master equation shows results that are qualitatively similar
to the classical solution in the regime of self-induced oscillations,
with the peaks being progressively broadened, reduced in height, and
shifted to lower detuning for increasing values of the quantum parameter
$\zeta$. Numerical evidence indicates that quantum correlations between
the cantilever position operator $\hat{x}$ and the photon operators
$\hat{a}^{\dagger},\,\hat{a}$ may cause the observed shift. As expected,
the discrepancy between the quantum mechanical and the classical result
reduces with diminishing quantum parameter $\zeta$. In Fig.~\ref{fig:4}(b),
we show the dependence of the cantilever energy on the quantum parameter,
for two different values of the detuning. In the sub-threshold regime
of amplification/heating the cantilever energy scales as $\zeta^{2}$,
as discussed above. In any case, the classical limit is clearly reached
as $\zeta\rightarrow0$. 

At the second sideband a classical solution of finite amplitude coexists
with a stable zero-amplitude solution (compare Fig.~\ref{fig:1}
and last panel of Fig.~\ref{fig:5}). The black curve in Fig.~\ref{fig:4}(b),
showing the finite amplitude solution, may therefore deviate substantially
from the $\hbar\rightarrow0$ limit of the quantum mechanical result.
In general, the average value of $E_{M}$, shown here, will be determined
by the relative weight of the two solutions (which are connected by
tunneling due to fluctuations), as well as fluctuations of $E_{M}$
for each of those two attractors.

\subsection{Langevin equation}

To get an estimate of the influence of quantum fluctuations, we compare
the results of the quantum master equation to numerical simulations
of classical Langevin equations that try to mimick the quantum noise.
The resulting description of the quantum-to-semiclassical crossover
is illustrated in Figs. \ref{fig:3}(c). To imitate both the zero-point
fluctuations of the mechanical oscillator and the shot-noise inside
the cavity, we add white noise terms to Eqs.\ref{eq:2}and \ref{eq:3}:\begin{equation}
\dot{\alpha}=[i(\Delta+g\frac{x}{x_{\text{ZPF}}})-\frac{\kappa}{2}]\,\alpha-i\alpha_{L}+\sqrt{\kappa/2}\,\alpha_{in}\label{eq:langevin_a}\end{equation}
\begin{equation}
\ddot{x}=-\omega_{M}^{2}x+\frac{\hbar g}{mx_{\text{ZPF}}}\left|\alpha\right|^{2}-\Gamma_{M}\dot{x}+\sqrt{\hbar\omega_{M}\Gamma/m}\,\xi,\label{eq:langevin_x}\end{equation}
\[
\]
where $\langle\alpha_{in}\rangle=\langle\xi\rangle=0$ and $\langle\alpha_{in}(t)\alpha_{in}^{*}(t')\rangle=\langle\xi(t)\xi(t')\rangle=\delta(t-t')$.
The coefficients in front of the noise terms are chosen such that
in the absence of optomechanical coupling we obtain the zero-point
fluctuations, i.e. $\left\langle |\alpha|^{2}\right\rangle =0.5$
away from resonance and $\frac{m\omega_{M}^{2}}{2}\langle x^{2}\rangle=\frac{\hbar\omega_{M}}{4}$.
The mean zero-point energy of the cantilever is substracted from the
curve in Fig.\ref{fig:3}(c). 

For parameters below the onset of self-sustained oscillations, this
semiclassical approach leads to good qualitative agreement with the
quantum mechanical description, as can be seen in Fig. \ref{fig:3}(c)
for parameters that are the same as those of \ref{fig:3}(a). Still,
the Langevin approach can mimick the results from the master equation
only partially. In particular, the approximation gets worse when dealing
with low photon numbers. This is because the Langevin equation introduces
artificial fluctuations of the radiation pressure force in the vacuum
state. Indeed, $\left|\alpha\right|^{2}$ has a finite variance even
in the ground state of the photon field, in contrast to $\hat{a}^{\dagger}\hat{a}$.

\subsection{Wigner density and phonon number distribution}

In figure~\ref{fig:5}, we go beyond the average cantilever phonon
number and present results both for the phonon number probability
distribution, as well as the full Wigner density of the cantilever,
defined as

\begin{equation}
W(x,p)=\frac{1}{\pi\hbar}\int_{-\infty}^{+\infty}\left\langle x-y\left|\hat{\rho}\right|x+y\right\rangle e^{2ipy/\hbar}\, dy.\end{equation}

\begin{figure}[H]
\begin{centering}
\includegraphics[width=0.95\columnwidth]{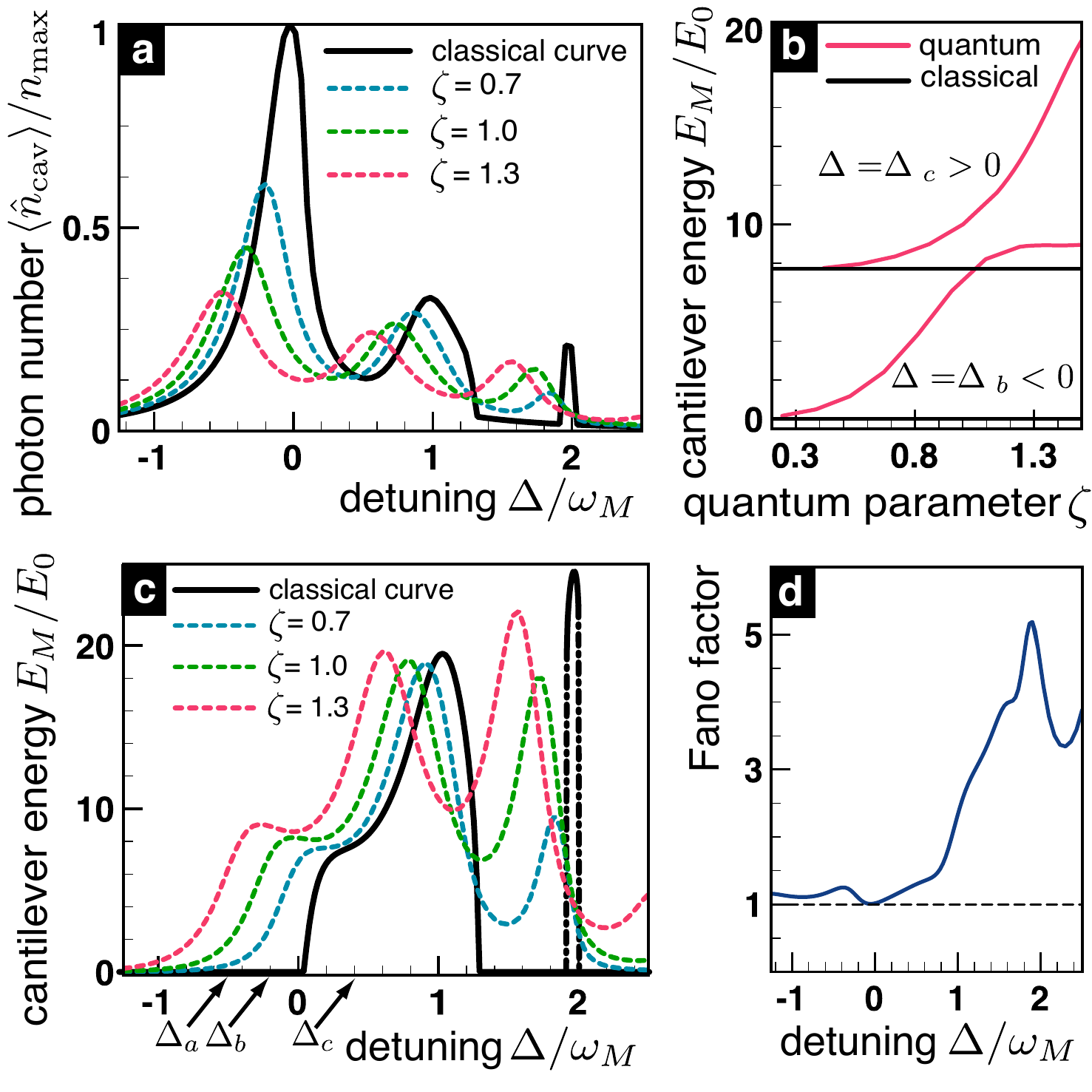}\caption{Comparison of classical and quantum results. (a) Number of photons
inside the cavity as a function of detuning, and (c) energy of the
cantilever versus detuning for $\Gamma_{M}^{*},\,{\cal P}^{*}$ and
$\kappa/\omega_{M}=0.5$. The dotted curves show results from the
quantum master equation for different values of the quantum parameter
$\zeta=1.3$ (pink) , $\zeta=1.0$ (green) and $\zeta=0.7$ (blue),
which are compared with the solution of the classical equations of
motion (black solid curve). As $\zeta\rightarrow0$, the qantum result
approaches the classical curve. See main text for a detailed discussion.
(b) The energy of the cantilever as a function of the quantum parameter
$\zeta$ for fixed detunings $\Delta_{b}/\omega_{M}=-0.2$ and $\Delta_{c}/\omega_{M}=0.4$
(the detuning value $\Delta_{a}$ indicated in (b) is used in Fig.~\ref{fig:5}).
(d) Fano factor $(\langle\hat{n}_{M}^{2}\rangle-\langle\hat{n}_{M}\rangle^{2})/\langle\hat{n}_{M}\rangle$
vs. detuning, for $\zeta=1$. For a coherent state whose occupation
number follows a Poisson distribution, the Fano factor is $1$ (dashed
black line). Close to the resonance (and far away from it, where $\left\langle \hat{n}_{M}\right\rangle =0$),
the results of the quantum master equation approach this value. The
Fano factor becomes particularly large near the second sideband, where
we observe coexistence of different oscillation amplitudes (see Fig.~\ref{fig:5}).
\label{fig:4}}

\par\end{centering}
\end{figure}

\begin{figure}[H]
\begin{centering}
\includegraphics[width=0.95\columnwidth]{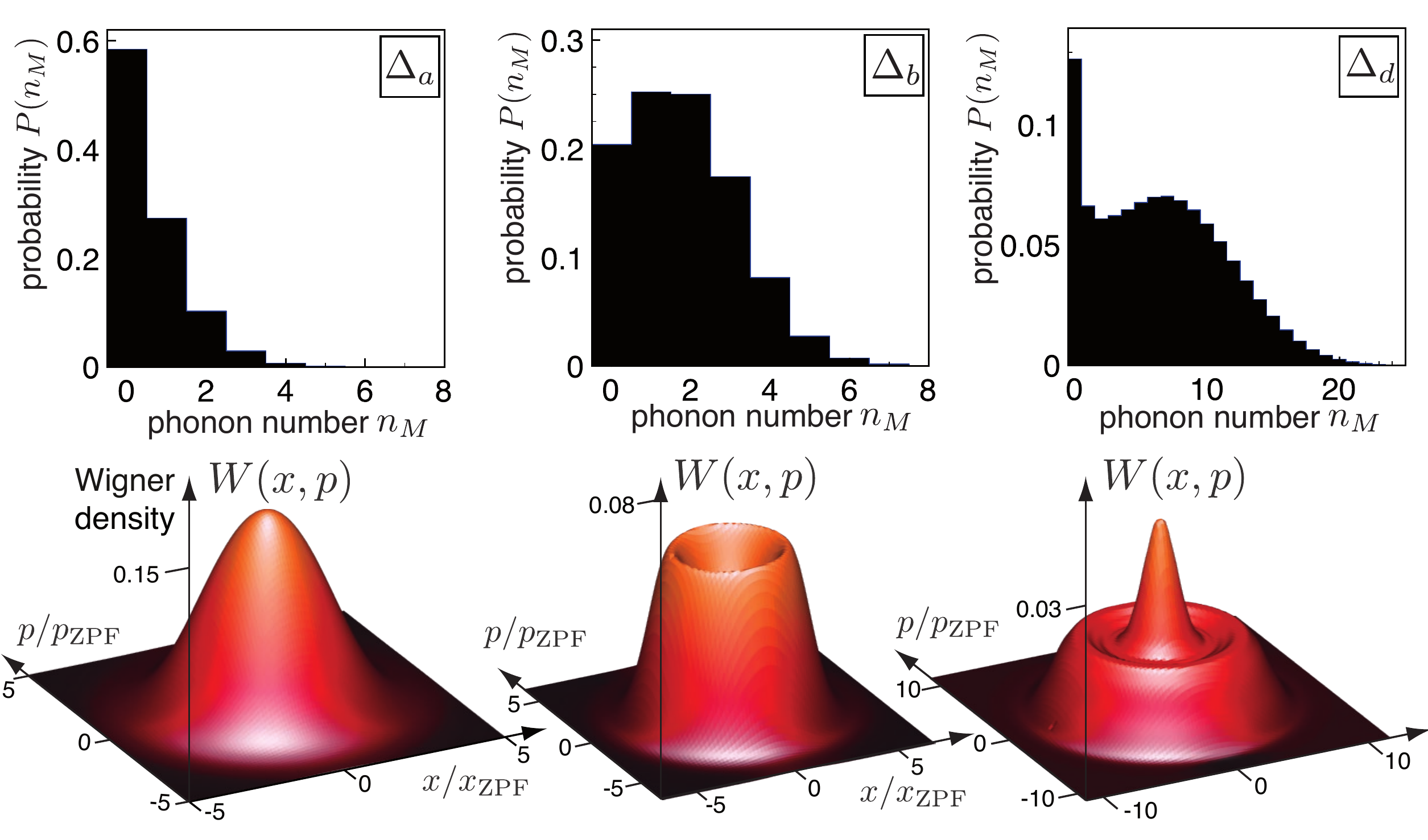}
\par\end{centering}

\caption{Distribution functions $P(n_{M})$ of the cantilever occupation and
Wigner functions $W(x,p)$ {[}rescaled by $x_{{\rm ZPF}}p_{{\rm ZPF}}$]
of the cantilever for $\Delta_{a}=-0.45\omega_{M}$, $\Delta_{b}=-0.2\omega_{M}$,
$\Delta_{d}=1.72\omega_{M}$ {[}corresponding to the detuning values
also indicated in Fig.~\ref{fig:4}(b); further parameters as in
Fig.~\ref{fig:4} with $\zeta=1.0$; for $\Delta_{d}$ the mechanical
damping rate is reduced to $\Gamma_{M}/\omega_{M}=1.2\cdot10^{-3}$].
Below the threshold of self-induced oscillations, a broadened distribution
is found corresponding to an increased effective temperature, cf.~Eq.~\eqref{eq:N_c_rate_equations}
(left panels, $\Delta_{a}$); self-induced oscillations are visible
as a finite amplitude ring in the middle and the right panel. Dynamical
multistability (i.e. co-existence of several attractors) in the classical
solution becomes apparent both in the distribution and the Wigner
density, where a double-peaked structure develops. \label{fig:5}}

\end{figure}

This figure demonstrates the different nature of the cantilever dynamics
in the sub-threshold regime and above threshold, where self-induced
oscillations occur. Below the threshold (for a detuning $\Delta_{a}=-0.45\omega_{M}$
as indicated in Fig.~\ref{fig:4}, quantum parameter $\zeta=1$,
and other parameters as in Fig.~\ref{fig:4}) the occupation of the
cantilever is thermal, with an effective temperature determined by
the effective optomechanical and mechanical damping rates, cf.~Eq.~\eqref{eq:N_c_rate_equations}.
Consequently, the Wigner density shows a broad peak around the origin
of the $x-p$ plane of cantilever position and momentum (the static
shift of the cantilever due to the radiation pressure is very small).
For a detuning of $\Delta_{b}=-0.2\omega_{M}$, self-induced oscillations
occur. The probability distribution for the phonon number shows some
thermal broadening, but an additional peak appears at a finite phonon
number. In the Wigner density plot this results in a crater-like feature,
which corresponds to a mixture of coherent states with essentially
fixed amplitude but arbitrary phases. This captures the fact that
the phase of the self-induced oscillations is completely arbitrary
also in the classical solution. The energy corresponding to the phonon
number at which the distribution peaks, compares fairly well to the
oscillation energy obtained from the classical solution. Only the
shift towards lower values of detuning as shown in Fig. \ref{fig:4}(b)
puts restrictions on a detailed quantitative comparison. %

For a value of the detuning located in the second sideband, $\Delta_{d}=1.72\omega_{M}$,
we find a probability distribution with a peak for the occupation
of the cantilever ground state, and a broader peak at a finite occupation
number (mechanical damping is slightly decreased to display more pronounced
features). Likewise, the Wigner density consists of a sharp peak at
the origin, surrounded by a broader ring representing finite amplitude
oscillations. This corresponds to the existence of two stable attractors
in the classical analysis, with vanishing and finite oscillation amplitude,
respectively. Similar results for the Wigner densities were found
in Ref.~\cite{2007_Armour_ResonatorSSET} for a cantilever driven
by a superconducting single-electron transistor.

\section{Conclusion}

We presented a fully quantum mechanical treatment of a driven optical
cavity coupled to a mechanical cantilever by radiation pressure. Light-induced
forces can yield a negative contribution to the damping of the cantilever,
causing amplification of fluctuations and even instabilities of the
cantilever dynamics. 

In the present paper we first reviewed briefly the classical solution
and discussed the existence of self-induced oscillations and the resulting
attractor diagram of the system. We paid particular attention to the
resolved-sideband regime $\kappa\ll\omega_{M}$, which is now increasingly
studied in experimental setups. Here the instabilities clearly occur
at sidebands, where the detuning matches an integer multiple of the
mechanical frequency.

Within a simple rate equation approach, we were able to discuss the
influence of the photon shot noise and quantum fluctuations well below
the instability threshold. The full quantum-mechanical treatment,
based on a numerical solution of the quantum master equation, is able
to completely describe both regimes (below and above threshold). It
has been complemented by numerical studies of a Langevin equation
that includes the zero-point fluctuations in a semiclassical way.
We studied the crossover between the quantum and classical regime,
which is governed by the quantum parameter, $\zeta=x_{\text{ZPF}}/x_{\text{FWHM}}\,,$
denoting the ratio between the mechanical zero-point fluctuation amplitude
and the width of the optical resonance. Signatures of the self-induced
oscillations are also found in the full quantum mechanical solution,
even at larger values of $\zeta$. In regions of dynamical multistability,
the different attractors show up simultaneously in the steady state
of the cantilever, since the quantum noise can induce transitions
between those attractors. Finally, we characterized the mechanical
motion in the various regimes by discussing the phonon number probability
distribution as well as the Wigner density.

\section*{Acknowledgments}

We thank A. Clerk, S. Girvin, K. Karrai, C. Neuenhahn, C. Metzger,
I. Favero, D. Rodrigues, and J. Harris for discussions and fruitful
collaboration on the optomechanical instability. We acknowledge support
by the DFG, in the form of the Nanosystems Initiative Munich (NIM),
the SFB 631, and the Emmy-Noether program. 

\bibliographystyle{unsrt}
\bibliography{BibFM}

\end{document}